\documentstyle[12pt]{article}
\begin{document}

\makeatletter
\def\@maketitle{\newpage
 \null
 {\normalsize \tt \begin{flushright} 
  \begin{tabular}[t]{l} \@date 
  \end{tabular}
 \end{flushright}}
 \begin{center}
 \vskip 2em
 {\LARGE \@title \par} \vskip 1.5em {\large \lineskip .5em
 \begin{tabular}[t]{c}\@author 
 \end{tabular}\par} 
 \end{center}
 \par
 \vskip 1.5em} 
\makeatother
\topmargin=-1cm
\oddsidemargin=1.5cm
\evensidemargin=-.0cm
\textwidth=15.5cm
\textheight=22cm
\setlength{\baselineskip}{16pt}
\title{ Conformal Invariance of the D-Particle Effective Action 
             }
\author{  Kiyoshi {\sc Hayasaka}\thanks{ hayasaka@particle.sci.hokudai.ac.jp}
        \  and Ryuichi {\sc Nakayama}\thanks{
                          nakayama@particle.sci.hokudai.ac.jp} 
\\[1cm]
{\small
    Department of Physics, Faculty of Science,} \\
{\small
           Hokkaido University, Sapporo 060-0810, Japan}
}
\date{
  EPHOU-99-005  \\
  hep-th/9904083 \\ 
  April   1999  
}
%
%
\maketitle

\begin{abstract}

It is shown that the effective theory of D-particles has conformal symmetry
with field-dependent parameters. This is a consequence of the supersymmetry. 
The string coupling constant is not transformed in contrast with the
recent proposal of generalized conformal symmtery 
by Jevicki et al.\cite{JY}\cite{JKY2}
This conformal symmetry can also be generalized to other Dp-brane systems.

\end{abstract}
\newpage
\newcommand {\beq}{\begin{equation}}
\newcommand {\eeq}{\end{equation}}
\newcommand {\beqa}{\begin{eqnarray}}
\newcommand {\eeqa} {\end{eqnarray}}
\newcommand{\bm}[1]{\mbox{\boldmath $#1$}}

\section{Introduction}
\hspace{5mm}

Duality between string theory in Anti-de Sitter (AdS) background space-time 
and conformal field theory was conjectured in \cite{Mal}. This correspondence
was further elaborated in \cite{GKPW}. Especially near-horizon geometry of 
nearly coincident D3-branes\cite{Dbrane} is $AdS_5 (\times S^5)$ and its 
isometry is the conformal symmetry. On the other hand four-dimensional 
${\cal N}=4$ supersymmetric Yang-Mills (SYM) theory is conformally invariant. 
It was shown in \cite{JKY1} that there is a correspondence between the 
conformal symmetries on both sides.

Near-horizon geometry of other Dp-branes is not AdS space-time and 
this conjecture does not simply apply.  The corresponding SYM theory is 
not apparently conformally invariant because the coupling constant is 
dimensionful. Recently, however, it was pointed out by Jevicki et 
al\cite{JY}\cite{JKY2} that the near-horizon geometry of Dp-branes can 
also be interpreted as conformally invariant, if one varies the string 
coupling constant together with other backgrounds.  Then they claimed 
that by regarding the coupling constant of SYM theory as a background
field and transforming this background appropriately with other fields, 
they can also make SYM theory confomally invariant.  They also argued 
that the 2 conformal transformations on the SYM and supergravity sides  
are related by some coordinate transformations.\cite{JKY2}

We will not adopt this proposal to regard the coupling constant as a 
background, because a coordinate-dependent coupling constant
breaks the special conformal symmetry as well as  supersymmetry (SUSY).
(See sec 4.) The coupling constant must be kept constant.  In this paper
we will specialize to the D-particle system and show that this system has 
conformal symmetry. For this purpose we will use the SUSY
transformation to induce a variation of the functional measure which is 
equivalent to effectively changing the coupling constant. Additional BRS 
transformation  needs to be also performed.  The ordinary conformal 
transformation combined with the SUSY and BRS transformation yields the 
desired, complete conformal transformation.

In sec 2 we will show that by introducing an auxiliarly field the SUSY 
transformation can be extended to an off-shell-closed symmetry, with respect 
to whose generator the D-particle action can be rewritten as an exact form. 
This is the T-dual version of the nil-potent symmetry studied in \cite{MNS}.
In sec 3 this symmetry is used to effectively change the string coupling 
constant in the action by transformation of the field variables. 
The parameter of the transformation is chosen to be field-dependent 
and non-local. In order to cancel the variations of the gauge-fixing and 
ghost action BRS transformation with a field-dependent parameter needs to 
be also performed. In sec 4 these results are used to show that D-particle 
effective theory has conformal symmetry. Discussions will be given in sec 5.

\section{Q-symmetry of the D-particle effective action}
\hspace{5mm}

The action which describes the low-energy dynamics of 
D-particles\cite{Dbrane} is given \cite{Eaction} by
\beq
S = \int dt \frac{1}{g} \mbox{Tr} \left( \frac{1}{2} \sum_{i=1}^9(D X^i)^2
+ \frac{1}{4} \sum_{i,j=1}^9 [X^i,X^j]^2 -\frac{i}{2} \psi^T D \psi 
- \frac{1}{2} \sum_{i=1}^9 \psi^T \gamma^i [X^i,\psi] \right),
\label{DPaction}
\eeq
where $D= \partial_t -i A^0$ is a gauge covariant derivative and $X^i$,
$A^0$ and $\psi$ are N $\times$ N hermitian matrices.  $\psi$ is also a 
sixteen-component spinor.  $g$ is the string coupling constant.
$T$ on $\psi$ stands for a transposition. Some conventions for $\gamma$ 
matrices and the spinor $\psi$ are given in Appendix.
In what follows summation symbols will be omitted. Indices $i,j, \ldots$
run from 1 to 9, and $a,b,\ldots$ from 1 to 8.

Action (\ref{DPaction}) is invariant under the extended SUSY 
transformation.\cite{BFSS}
\beqa
\delta X^i & = &-i \epsilon^T \gamma^i \psi, \qquad
\delta A^0  =  i \epsilon^T \psi, \nonumber \\
\delta \psi & = & - (DX^i) \gamma^i \epsilon - \frac{i}{2} [X^i,X^j]
\gamma^{ij} \epsilon + \epsilon' 
\label{Nsusy}
\eeqa
Let us pick up a particular transformation with $\epsilon'=0$, 
$\epsilon = \varepsilon \zeta$ and $\zeta= (\frac{1}{\sqrt{2}},0,0,
\frac{1}{\sqrt{2}},0,\cdots,0)^T$. $\varepsilon$ is a Grassmann odd constant. 
Generality will not be lost by this choice of $\epsilon$, because any 
$\epsilon$ can be put in this form by SO(9) rotation. 
By substituting the representation of $\gamma$'s (\ref{A2}), (\ref{A3})
and that of $\psi$ (\ref{B1}), (\ref{B2}) into (\ref{Nsusy}), we explicitly 
obtain the SUSY transformation of the component fields. 
\beqa
\delta X^a  & = & -i \varepsilon \psi_a \quad (a=1,\cdots,8),  \qquad
\qquad 
\delta X^9  = -i \varepsilon \eta, \nonumber \\
\delta A^0 & = & i \varepsilon \eta, \qquad \qquad \qquad   
\delta \psi_a  = i \varepsilon ([X^9,X^a] +i DX^a) \quad (a=1,\cdots,8), 
                         \nonumber \\
\delta \vec{\chi}  & = & -\frac{1}{2} \varepsilon \vec{ E},  \qquad
\qquad 
\delta \eta = - \varepsilon D X^9 
\label{susy}
\eeqa
Here $\psi_a$, $\vec{\chi}=(\chi^1,\cdots,\chi^7)$ and $\eta$ are components 
of $\psi$ and $\vec{E}$ a seven-vector function 
of $X^a$. These are defined  in Appendix.

We could achieve the goal of this paper by working directly with 
transformation (\ref{susy}). It is, however, more instructive to  relate it  
to the nil-potent transformation of \cite{MNS}.  
Let us  introduce an auxiliarly seven-vector
$\vec{H}$ by adding a term $\int dt (2g)^{-1} \mbox{Tr} (\vec{H}+\frac{1}{2} 
\vec{E})^2$ to (\ref{DPaction}). $\vec{H}$ will coincide with $-\frac{1}{2} 
\vec{E}$ by the equations of motion. The transformation rule of $\vec{\chi}$ 
is modified to $\delta \vec{\chi} = \varepsilon \vec{H}$. $\delta \vec{H} $ is 
defined to be $\varepsilon ([X^9,\vec{\chi}]+ iD \vec{\chi})$. 
Now the square of this new  
transformation turns out to be equivalent to a time translation up to a gauge
transformation: $ [Q^2, \vec{\chi} ] = i\partial_t \vec{\chi} 
+ [X^9+A^0, \vec{\chi} ]$, {\it etc}.\footnote{
The algebra of transformation (\ref{Nsusy})  closes only on shell, {\it i.e.}
when equations of motion are used. By the introduction of $\vec{H}$, 
Q part of the SUSY algebra closes off shell.}
Here $Q$ is the generator of this 
transformation.($\delta=\varepsilon Q$) We will hereafter call this 
a Q transformation.  Importantly, action (\ref{DPaction}) can be 
rewritten into a Q-exact form.   
\beq
S_{coh} = \int dt \frac{1}{g} \left\{ Q, \mbox{Tr} \left( 
-\frac{1}{2} \eta DX^9 + 
\frac{1}{2} \vec{\chi} \cdot \vec{E} + \frac{1}{2} \vec{\chi} \cdot
\vec{H} + \frac{i}{2} \psi_a [X^a,X^9] - \frac{1}{2} \psi_a DX^a 
\right) \right\}
\label{Cohaction}
\eeq
This action is the T-dual version of the cohomological action for D-instantons
considered in \cite{MNS}.

To quantize this model we will choose the background field gauge.
We will decompose $X^i$ 
into a background field $B^i$ and a quantum fluctuation $Y^i$
\beq
X^i = B^i + Y^i
\eeq
and choose a gauge function\cite{JKY2}
\beq
G= \partial_t A^0 + i [ B^i, Y^i ] .
\eeq
The gauge fixing and  ghost actions are given by
\beqa
S_{gf} & = & \int dt \ \frac{-1}{2g} \ \mbox{Tr} G^2, \\
S_{gh} & = & \int dt \ i \ \mbox{Tr} \left\{ \bar{C} \left( \partial_t D C 
+ [ B^i,[X^i,C ] \ ] \right) \right\} .
\eeqa
The total action $S_{tot} = S_{coh}+S_{gf}+S_{gh}$ is invariant under the
BRS transformation  
\beqa
\delta_B X^i & = & -\lambda \ [C,X^i], \qquad \qquad 
\delta_B A^0  =  i \ \lambda \ DC, \nonumber \\
\delta_B \psi & = & - \lambda \ \{ C, \psi \}, \qquad \qquad 
\delta_B \vec{H}  =  - \lambda \  [C,\vec{H}], \nonumber \\
\delta_B C & = & - \lambda \ C^2, \qquad \qquad 
\delta_B \bar{C}  =  \lambda \  \frac{1}{g} \ G.
\label{BRS}
\eeqa
Here $\lambda$ is a Grassmann odd parameter.

\section{Transformation with a field-dependent parameter}
\hspace{5mm}
Because of the Q-exact nature of action (\ref{Cohaction}) the
string coupling constant $g$ in (\ref{Cohaction})  can be effectively 
changed by a Q transformation of fields without varying $g$ explicitly. 
The variation $\delta g$ may even be a
function of $t$. We will carry out this program in two steps.

We first perform a  Q transformation 
\beqa
\delta_Q X^a  & = & -i \varepsilon \psi_a,  \qquad
\qquad 
\delta_Q X^9  = -i \varepsilon \eta, \nonumber \\
\delta_Q A^0 & = & i \varepsilon \eta, \qquad \qquad \qquad   
\delta_Q \psi_a  = i \varepsilon ([X^9,X^a] +i DX^a), 
                         \nonumber \\
\delta_Q \vec{\chi}  & = & \varepsilon \vec{H},  \qquad
\qquad 
\delta_Q \eta = - \varepsilon D X^9, \qquad 
\delta_Q \vec{H}  =  \varepsilon ([X^9, \vec{\chi}] + i D \vec{\chi}), 
                         \nonumber \\ 
\delta_Q C & = & 0, \qquad \delta_Q \bar{C} =0,
\label{Qtransf}
\eeqa
with the following parameter.
\beq
\varepsilon = 
\int dt \frac{i}{g^2} \delta g(t) \mbox{Tr} \left(  
-\frac{1}{2} \eta DX^9 - \frac{1}{2} \psi_a DX^a + 
\frac{1}{2} \vec{\chi} \cdot \vec{E} + \frac{1}{2} \vec{\chi} \cdot
\vec{H} + \frac{i}{2} \psi_a [X^a,X^9] 
\right) 
\label{epsilon}
\eeq
This is a field-dependent and non-local transformation.  
Action(\ref{Cohaction}) is invariant,
while the gauge fixing and ghost actions are changed.
\beqa
\delta_Q  S_{gf} & = & -\varepsilon \int dt \frac{1}{g} \mbox{Tr} \{ G
\left(i \partial_t \eta +[B^a,\psi_a ] +[B^9,\eta]
\right)   \}, 
\label{varSgf} \\
\delta_Q  S_{gh} & = & -\varepsilon \int dt \mbox{Tr} \{ \bar{C} \left(
i \partial_t \{ \eta,C \} +[B^a, \{ \psi_a,C \} \ ] + [ B^9, \{ \eta,C
\} \ ] 
\right) \} \qquad \qquad 
\label{varSgh}
\eeqa
The functional measure is not invariant, either, because the transformation 
parameter depends on the fields.  Calculation of the superjacobian is 
straightforward and we obtain\footnote{
Similar calculation of the superjacobian for 4d bosonic gauge theory 
in a different context was performed in \cite{FradPal}. See also \cite{JKY2}.}

\vspace{.3cm}

$(1+ \delta_Q) {\cal D} (\mbox{Fields}) =$ \\ 

\vspace{.1cm}

$\displaystyle \ \ \  {\cal D} (\mbox{Fields})
\exp \left\{  i \int dt \frac{-\delta g(t)}{g^2} \mbox{Tr} \left( -\frac{i}{2}
\psi_a D \psi_a + \frac{1}{2} \psi_a [X^9,\psi_a] - \frac{1}{2} \vec{\chi} 
\cdot [Q, \vec{E}]
\right. \right. $
\beqa
\qquad \qquad &  & - \frac{1}{2} \vec{\chi} \cdot [X^9, \vec{\chi}] 
- \frac{i}{2} \vec{\chi} \cdot D \vec{\chi} + \frac{1}{2} [X^9, X^a]^2 
+ \frac{1}{2} (DX^a)^2 + \frac{1}{2} (DX^9)^2 \nonumber \\
 \qquad \qquad &  & \left. \left. -\frac{i}{2} \eta D \eta 
- \eta [X^a,\psi_a] - \frac{1}{2} \eta [X^9,\eta]
 + \frac{1}{2} \vec{H}^2 + \frac{1}{2} \vec{H} \cdot \vec{E} 
  \right)   \right\} . 
\label{varMeasure}
\eeqa
The exponent on RHS is equal to the difference $i \{ S_{coh}(g+\delta g) 
-S_{coh}(g)\}$. Because $S_{gh}$ does not depend on $g$, and the variation of
$S_{gf}$ with respect to $g$ is BRS exact and does not contribute to the path 
integral, the effect of the transformation (\ref{Qtransf}) is just to change 
$g$ to $g+\delta g(t)$ in $S_{tot}$.

Secondly, to cancel the variations (\ref{varSgf}) and  (\ref{varSgh}) we 
perform the BRS transformation (\ref{BRS}) with the parameter 
\beq
\lambda = \varepsilon \int dt \mbox{Tr} \left\{ \left( \partial_t \eta -i 
[B^a,\psi_a] -i [B^9,\eta] \right) \ \bar{C} \right\} .
\label{lambda}
\eeq
While the total action is invariant, the functional measure changes as 
in (\ref{varMeasure}).  It can be checked that this change  cancels out 
(\ref{varSgf}) and (\ref{varSgh}) exactly.
 
To summarize, the combination 
\beq
\delta=\delta_Q+\delta_B
\label{varQB}
\eeq
has the same effect as changing $g$ into $g+\delta g(t)$. It is important to
notice that this infinitesimal transformation cannot be repeated to generate
a non-constant $g(t)$, because Q symmetry is broken if $g$ is not a constant. 

It is straightforward to eliminate $\vec{H}$ from (\ref{varQB}) by using
$\langle \vec{H}\rangle = -\frac{1}{2} \vec{E}$ and $\langle H_{\alpha 
\beta}^A(t) H_{\gamma \delta}
^B(t') \rangle = ig \delta^{AB} \delta_{\alpha \delta} \delta_{\beta \gamma} 
\delta (t-t') + \frac{1}{4} E_{\alpha \beta}^A(t) E_{\gamma \delta}^B(t')$. 
Here $\alpha,\beta,\gamma,\delta=1,\cdots,$N and $A, B=1,\cdots,7$ are 
U(N) and vector indices, respectively.  
The results are the same as (\ref{varQB}) except for replacement of 
$\varepsilon$ (\ref{epsilon}) and $\lambda$ (\ref{lambda}) by
\beqa
\tilde{\varepsilon} & = & 
\int dt \frac{i}{g^2} \delta g(t) \mbox{Tr} \left(  
-\frac{1}{2} \eta DX^9 - \frac{1}{2} \psi_a DX^a
+\frac{1}{4} \vec{\chi} \cdot \vec{E}  + \frac{i}{2} \psi_a [X^a,X^9] 
\right) \nonumber \\
& = &\int dt \frac{i}{g^2} \delta g(t) \mbox{Tr} \left\{ \zeta^T \left(  
-\frac{1}{2} \gamma^i DX^i - \frac{i}{4} \gamma^{ij} [X^i,X^j] \right) \psi 
\right\}                           \label{tilep} 
\eeqa
and 
\beqa
\tilde{\lambda} & = & 
\tilde{\varepsilon} \int dt \mbox{Tr} \left\{ \left( \partial_t \eta -i 
[B^a,\psi_a] -i [B^9,\eta] \right) \, \bar{C} \right\} \nonumber \\
& = & \tilde{\varepsilon} \int dt \mbox{Tr} \left\{ \zeta^T \left( \partial_t
\psi - i \gamma^i [B^i,\psi] \right) \, \bar{C} \right\},
                                \label{tillam}
\eeqa
respectively, and 
\beq
\delta \vec{\chi} = -\frac{1}{2} \tilde{\varepsilon} \vec{E} - \tilde{\lambda}
\{C,\vec{\chi} \} -\frac{\delta g(t)}{2g} \vec{\chi},
\eeq
where the last term comes from the contraction of 2 $\vec{H}$'s.

\section{Conformal symmetry}
\hspace{5mm}

It was claimed in \cite{JKY2} that Dp-brane action has conformal symmetry 
provided one regards $g$ as a background field $g(x)$ and makes it 
transformed appropriately together with other fields.  They called this 
symmetry a generalized conformal one.  

The conformal group is generated by translation, dilatation and 
special conformal transformation (SCT).  We will specialize to the
D-particle case. These 3 transformations on  $t$  are defined by
\beqa
\delta t & = &  -a \quad (\mbox{translation}), \\
\delta t & = & -a t \quad (\mbox{dilatation}), \\
\delta t & = & -a t^2 \quad (\mbox{SCT}). 
\eeqa
Here $a$ is an infinitesimal parameter. A field $F(t)$ of a scale dimension 
$w$ is then transformed as 
\beq
\delta_n F(t) = a (nwt^{n-1} + t^n \partial_t) F(t) 
\eeq
for translation ($n=0$), dilatation ($n=1$) and SCT ($n=2$).

Actions $S$ and $S_{coh}$ are invariant under these transformations only if
the string coupling constant $g$ is also transformed like
\beq
\delta_n g = 3 a n t^{n-1} g.
\label{varg}
\eeq
The scale dimensions $w$ of the fields are  1, 1, $\frac{3}{2}$, 2
for $X^i$, $A^0$, $\psi$ and $\vec{H}$, respectively.
Because (\ref{varg}) for SCT ($n=2$) depends on $t$ explicitly, the authors of
\cite{JY} \cite{JKY2} proposed to regard $g$ itself as a function of 
the coordinate $t$ and assumed a transformation rule
\beq
\delta_n g(t) = a(3nt^{n-1}+ t^n \partial_t) g(t).
\eeq
As for $S_{gf}$ and $S_{gh}$ these are not invariant under SCT
even if $C$ and $\bar{C}$ are assigned scale dimensions 0 and $-1$.
By using the formalism  of \cite{FradPal} the authors of \cite{JKY2} then
 showed that the variations of these actions 
can be cancelled by a non-local BRS transformation (\ref{BRS})   
with $\lambda = \nu_n$.\footnote{Here $n=2$ for SCT. An index $n$ is 
introduced for later convenience.          }
\beq
\nu_n = -n(n-1)ia \int dt \mbox{Tr} \bar{C} A^0
\label{BRSnu}
\eeq

We do not adopt their proporsal to regard $g$ as a $t$-dependent background 
field $g(t)$ because of the following reasons.

\begin{itemize}
\item Once $g$ is promoted to a function of $t$,  then action (\ref{DPaction}) 
is {\em no longer} invariant under SCT because the integration by parts 
is used in the proof of invariance. 
After a simple calculation we indeed obtain
\beq 
\delta_2 S = a \int dt \frac{1}{g(t)} \partial_t \mbox{Tr} (X^i)^2.
\label{vS}
\eeq
\item If  $g$ is a function of $t$, $S$ is {\em not} invariant under 
SUSY transformation (\ref{Nsusy}) due to the same reason as above. 
This symmetry is important and cannot be abandoned. 
\end{itemize}

For these reasons we will leave $g$ a constant and will not change it.
In the previous sections we showed that the same effect as changing $g$
can be accomplished by transforming the field variables. 
By combining (\ref{varQB}) and the BRS transformation (\ref{BRS}) with 
the parameter (\ref{BRSnu}), we obtain the 3 conformal transformations 
($n=0,1,2$)\footnote{In fact we do not change $g$.  Instead we use 
transformation (\ref{varQB}) to cancel the effective variation  of $g$ due to 
the conformal transformation. Thus $g$ remains constant. This also enables 
us to repeat infinitesimal transformations to obtain a finite one.},
\beqa
\Delta_n X^i & = & a(nt^{n-1}+t^n \partial_t)X^i -i \tilde{\varepsilon}_n 
\zeta^T \gamma^i \psi-(\tilde{\lambda}_n+\nu_n) [C,X^i],   \nonumber \\\Delta_n A^0 & = & a(nt^{n-1}+t^n \partial_t)A^0 + i \tilde{\varepsilon}_n
\zeta^T \psi+ i (\tilde{\lambda}_n+\nu_n)DC,  \nonumber \\
\Delta_n \psi_a & = & a \left(\frac{3}{2}nt^{n-1} +t^n \partial_t \right) 
\psi_a +i \tilde{\varepsilon}_n ([X^9,X^a]+iDX^a) -(\tilde{\lambda}_n+\nu_n)
 \{C,\psi_a \},             \nonumber \\
\Delta_n \vec{\chi} & = & a \left(\frac{3}{2}nt^{n-1} +t^n \partial_t \right)
     \vec{\chi}-\frac{1}{2} \tilde{\varepsilon}_n \vec{E} -
\frac{\delta_ng}{2g} \vec{\chi}-(\tilde{\lambda}_n+\nu_n) \{C,\vec{\chi} \}, 
                    \nonumber \\
\Delta_n \eta & = & a \left(\frac{3}{2}nt^{n-1} +t^n \partial_t \right) \eta
 -\tilde{\varepsilon}_n DX^9-(\tilde{\lambda}_n+\nu_n) \{C,\eta \},
              \nonumber \\
\Delta_n C & = & at^n \partial_t C -(\tilde{\lambda}_n+\nu_n)C^2,
            \nonumber \\
\Delta_n \bar{C} & = & a(-nt^{n-1} +t^n \partial_t)\bar{C} 
+(\tilde{\lambda}_n+\nu_n) \frac{1}{g} G.
\label{Conformal}
\eeqa
Here the first 2 lines are simplified by using $\zeta$.
$\tilde{\varepsilon}_n$ and $\tilde{\lambda}_n$ are (\ref{tilep}) and
(\ref{tillam}), respectively,  
with $\delta g$ replaced by $\delta_ng$
(\ref{varg}).  

In the rest of this section we will compute the expectation value 
$\langle \Delta_n X^i \rangle$, which will become a symmetry 
transformation $\Delta_n B^i$ of the effective action $\Gamma [B^i]$.  
This may be performed in perturbative expansions in $g$.
Because $\tilde{\varepsilon}_n$ (\ref{tilep}) and $\tilde{\lambda}_n$
(\ref{tillam}) are of order ${\cal O} (g^{-1})$, it turns out that 
in order to obtain ${\cal O} (g^m)$ result we have to perform $(m+1)$-loop 
calculation. Here we will present only the result of one-loop calculation 
( ${\cal O} (g^0)$). The result and details of two-loop calculation will be 
reported in a forthcoming paper \cite{HN}. 

The background field $B^i(t)$ is diagonal and linear in $t$.  
We found up to first order in $\dot{B}=\partial_t B$ that
$\langle -\tilde{\lambda}_n [C,X^i] \rangle$ is of order ${\cal O} (g^1)$ and 
\beq
\langle-i \tilde{\varepsilon}_n \zeta^T \gamma^i \psi \rangle_{\alpha \beta}  
=  -\frac{1}{2} \delta_{\alpha \beta} \int dt' \frac{\delta_n g(t')}{g} \theta
(t-t')  \dot{B}_{\alpha}^i(t').
\eeq 
Here $B^i_{\alpha}$ is the $\alpha$-th diagonal component of $B^i$ and 
$\theta (t-t')$ is a step function. $\langle-\nu_n [C,X^i] \rangle$ 
was obtained in \cite{JKY2} and is also of order 
${\cal O} (g^1)$. We thus obtain the \lq quantum' transformation rule
\beq
\Delta_1 B^i_{\alpha}(t) =  a(1+t \partial_t) B^i_{\alpha}(t) 
-\frac{3}{2} a B^i_{\alpha}(t) = a(-\frac{1}{2} +t \partial_t) B^i_{\alpha}(t)
\label{dil}
\eeq
for dilatation and
\beqa
\Delta_2 B^i_{\alpha}(t) & = & a(2t + t^2 \partial_t) B^i_{\alpha}(t) 
- 3a \int^t dt' \ t' \ \dot{B}^i_{\alpha}(t') \nonumber \\
& = & a(-t + t^2 \partial_t) B^i_{\alpha}(t) + 3a \int^t dt' B^i_{\alpha}(t')
\label{sct}
\eeqa
for SCT.  It is easy to verify that the tree-level effective action
\beq
\Gamma_0 [B] = \int dt \frac{1}{2g} (\dot{B}^i)^2
\label{Eaction}
\eeq
is left invariant under (\ref{dil}) and (\ref{sct}).
We note that the scale dimension $w$ of the backgound $B^i_{\alpha}$ 
has shifted from 1 to $-\frac{1}{2}$ due to quantum effects. 
We also find that SCT rule (\ref{sct}) acquired an extra non-local term. 
Nonetheless we can check that (\ref{dil}), (\ref{sct}) and 
$\Delta_0 B^i_{\alpha} =a \partial_t B^i_{\alpha}$ generate the conformal 
algebra.

\section{Discussion}
\hspace{5mm}

We found that the low-energy effective theory of D-particles is 
invariant under conformal transformation (\ref{Conformal}).
The transformation rule is obtained explicitly, although the parameters 
depend on the field variables in a non-local way.  
Here we  stress the point that $g$ is {\em not} changed under the 
transformation. We also found that the scale dimension of $X^i$ changed from 
1 to $-\frac{1}{2}$ due to quantum effects.

Because the actions of Dp-branes are related by T-dual transformation
\cite{Taylor}, the procedure adopted in this paper is easily generalized 
to all Dp-branes.
It can be verified that all Dp-brane theories have similar 
conformal symmetry. This conformal symmetry will put strong constraints on 
the correlation functions. Some lower-point functions may be obtained by 
solving conformal Ward-Takahashi identities.\cite{FradPal}  
Action (\ref{DPaction}) also defines the Matrix 
theory \cite{BFSS}. Study of the conformal W-T identities may also shed some 
light on the understanding of M theory. 

Another important issue is the AdS/CFT correspondence.
Because the near-horizon geometry of Dp-branes ($p \neq 3$) is 
not an AdS space-time and the conformal group is not its isometry,
this correspondence does not simply apply.  Because Dp-brane
system has turned out to have conformal symmetry, however, it is expected that 
its near-horizon geometry may also have the corresponding \lq conformal
symmetry'. If such a symmetry is found, Dp-brane effective theory may also be 
interpreted as a boundary \lq conformal field theory' in the classical 
background of the Dp-branes. It will play an important r\^{o}le in 
determining the effective action for the radial distance of a probe Dp-brane 
in the background field of N coincident Dp-branes placed at the 
origin.\cite{Mal} The non-local nature of transformation (\ref{sct}),
however, makes the problem difficult and the realization of such symmetry 
on the supergravity side is not yet clear.
The calculation of $\langle \Delta_n X^i \rangle$ to higher orders may 
elucidate this point. The result will be reported elsewhere\cite{HN}.
 
Finally, generalization of the present analysis to the superconformal 
transformation will be straightforward and may be useful in putting further 
constraints on the Dp-brane dynamics.



\section*{Note added}
\hspace{5mm}
If $g(t)$ is set a constant after SCT, the variation of $S$ (\ref{vS}) 
vanishes.  We were informed by T.~Yoneya that this is the generalized 
conformal symmetry of \cite{JY}\cite{JKY2}. We were also informed that he 
could extend this conformal symmetry to that for a non-constant $g$ by 
using the SO(2,1) orbit of $g$.  We thank T.~Yoneya for comments and 
discussions.

\section*{Appendix}
\hspace{5mm}
$\gamma^i, \quad i=1, \cdots, 9$ are 16 $\times$ 16 real 
symmetric matrices and satisfy Clifford algebra
\beq
\{\gamma^i, \gamma^j \}= 2 \delta^{ij}.
\eeq
We choose the following special representation.
\beqa
\gamma^i &  = & i \sigma_2 \otimes \mu^i = \left( \begin{array}{cc}
                    0 & \mu^i \\
                 - \mu^i & 0 \end{array} 
\right) \quad  (i=1,\cdots,7), \nonumber \\
\gamma^8  & = & \sigma_1 \otimes {\bm 1}_8 = \left( \begin{array}{cc}
                  0 & {\bm 1}_8 \\
                {\bm 1}_8  & 0 \end{array} \right), \nonumber \\
\gamma^9 &  = & \sigma_3 \otimes {\bm 1}_8 = \left( \begin{array}{cc}
                  {\bm 1}_8 & 0 \\
                   0 & - {\bm 1}_8 \end{array} \right) 
\label{A2}
\eeqa
8 $\times$ 8 matrices $\mu^i$ are given by 
\beqa
\mu^1 & = & i \sigma_2 \otimes i \sigma_2 \otimes i \sigma_2,  \qquad 
\mu^2 = {\bm 1}_2 \otimes  \sigma_1 \otimes i \sigma_2,  \qquad 
\mu^3  = {\bm 1}_2 \otimes \sigma_3 \otimes i \sigma_2, \nonumber \\
\mu^4 & = &  \sigma_1 \otimes i \sigma_2 \otimes {\bm 1}_2, \qquad 
\mu^5 =  \sigma_3 \otimes i \sigma_2 \otimes {\bm 1}_2,  \qquad 
\mu^6   =  i \sigma_2 \otimes {\bm 1}_2 \otimes \sigma_1, \nonumber \\
\mu^7 & = & i \sigma_2 \otimes {\bm 1}_2 \otimes \sigma_3. 
\label{A3}
\eeqa

The spinor $\psi$ is decomposed into 2 eight-component spinors $\psi^{(i)}$:
\beq
\psi = \left( \begin{array}{c}
                   \psi^{(1)} \\
                   \psi^{(2)} \end{array}      \right)
\label{B1}
\eeq
Their components are given by  
\beqa
\psi^{(1)} & = & \frac{1}{\sqrt{2}} (\eta+ \chi^7,
-\chi^2-\chi^4,\chi^2-\chi^4,\eta-\chi^7,\chi^1-\chi^6,-\chi^3-\chi^5,
-\chi^3+\chi^5,-\chi^1-\chi^6)^T, \nonumber \\
\psi^{(2)}  & = & \frac{1}{\sqrt{2}} (-\psi_2+\psi_8, \psi_3-\psi_5,
\psi_3+\psi_5,\psi_2+\psi_8,\psi_1+\psi_7,-\psi_4+\psi_6,\psi_4+\psi_6,
\psi_1-\psi_7)^T \nonumber \\
& & 
\label{B2}  
\eeqa

The seven-vector $\vec{E}=(E^1,\cdots,E^7)$ is defined by 
\beqa
E^1 & = & 2i \{ -[X^1,X^2]-[X^3,X^4]-[X^5,X^6]-[X^7,X^8] \}, \nonumber \\
E^2 & = & 2i \{ [X^1,X^4]+[X^2,X^3]-[X^5,X^8]-[X^6,X^7] \}, \nonumber \\
E^3 & = & 2i \{ -[X^1,X^3]+[X^2,X^4]-[X^5,X^7]+[X^6,X^8] \}, \nonumber \\
E^4 & = & 2i \{ [X^1,X^6]+[X^2,X^5]+[X^3,X^8]+[X^4,X^7] \}, \nonumber \\
E^5 & = & 2i \{ -[X^1,X^5]+[X^2,X^6]+[X^3,X^7]-[X^4,X^8] \}, \nonumber \\
E^6 & = & 2i \{ [X^1,X^8]+[X^2,X^7]-[X^3,X^6]-[X^4,X^5] \}, \nonumber \\
E^7 & = & 2i \{ -[X^1,X^7]+[X^2,X^8]-[X^3,X^5]+[X^4,X^6] \}. 
\label{C1}
\eeqa

\end{document}